\documentclass[aps,prl,twocolumn,epsfig,showpacs]{revtex4}
   
\usepackage{dcolumn}
\usepackage{bm}
\usepackage{graphicx}
\usepackage{amsmath}
\usepackage{latexsym}
\usepackage{amsfonts}
\usepackage{amssymb}
\usepackage{array}
\usepackage{epsfig}

\newcommand{\ket}[1]{\left\vert#1\right\rangle}
\newcommand{\bra}[1]{\left\langle#1\right\vert}

\begin{document}

\title{
Quantification of Macroscopic Quantum Superpositions within Phase Space
 }
\author{Chang-Woo Lee 
 and Hyunseok Jeong } 
\affiliation{
Center for Macroscopic Quantum Control \& Department of Physics and Astronomy,
Seoul National University, Seoul, 151-747, Korea
}
\date{\today}

\begin{abstract}
Based on phase-space structures of quantum states,
we propose a novel measure to quantify macroscopic quantum superpositions. Our measure simultaneously quantifies two different kinds of essential 
information for a given quantum state in a harmonious manner: the degree of quantum coherence and the effective size of the physical system that 
involves the superposition. It enjoys remarkably good analytical and algebraic properties. It turns out to be the most general and inclusive measure 
ever proposed that it can be applied to any types of multipartite states and mixed states represented in phase space.
\end{abstract}

\maketitle

\setlength\arraycolsep{1pt}

Quantum superposition is often considered the most
crucial feature of quantum mechanics.
Its evidence has been witnessed in numerous experiments using microscopic physical systems.
However, the question of whether
a truly macroscopic system could ever be in a quantum superposition
involves  far more nontrivial issues in both practical and philosophical aspects \cite{CatParadox}.
A macroscopic quantum superposition is supposed to consist of 
two (or more) macroscopically distinct states
but still maintains certain potential to manifest
quantum interference between the distinct component states.

Regarding the implementation of  macroscopic quantum superpositions,
limited but interesting progress
has been made in 
atomic/molecular systems \cite{MonroeCat,C60},
superconducting circuits \cite{SQUID1,SQUID2}, and
optical setups \cite{Ourjoumtsev,Gao,Afek}.
In particular, superpositions of coherent states (SCSs)
\cite{Ourjoumtsev}, multi-mode Greenberger-Horne-Zeilinger (GHZ) states \cite{Gao},
and NOON states \cite{Afek} have been experimentally demonstrated
in optical systems. 
Interestingly,
certain types of ``bigger-size'' superpositions 
may be useful for 
quantum information processing
\cite{CatInfo}.
However, even though various types of macroscopic superpositions 
have been theoretically studied and some experimental success made, 
the definition of their measure applicable to all those states
has remained a difficult yet urgent task.

Attempts to find a good measure for macroscopic quantum 
superpositions date back to Leggett \cite{Leggett}.
It has been followed by several
proposals \cite{Dur,Bjork,Korsbakken,Mar,Cavalcanti} and
each of them has its own merit and insight.
In those proposals, people often start from considering
the effective number of particles that involve the superposition
\cite{Leggett,Dur,Korsbakken}.
It could also be natural to take notice of distance
between the component states 
\cite{Dur,Bjork,Korsbakken,Cavalcanti,Mar}. 
Typically, these measures also depend on the choice of a specific target state \cite{Dur,Mar},
or a decomposition/observable \cite{Bjork,Cavalcanti,Korsbakken}.  

First of all,
it is crucial to note that
the number of effective particles (or distance between the component states)
cannot witness the true quantum superposition when determining the size of a macroscopic superposition.
These factors do not allow one to conclusively discriminate between a coherent
superposition 
and a classical mixture,
not to mention partially mixed states.
This problem was also pointed out in Ref.~\cite{Cavalcanti} where a signature of a
macroscopic superposition was studied.
This is unignorable because macroscopic superpositions 
typically lose quantum coherence, at least to some extent,
due to interactions with their environments, which process is called decoherence \cite{deco}.
In other words, a proper measure for a macroscopic superposition
must quantify the degree of a true superposition against an incoherent mixture,
together with its effective size factor such 
as the effective number of particles.

Furthermore, 
it should be pointed out that
the choice of a target state or of a
fiducial decomposition/observable, which the aforementioned measures employ,  
is actually arbitrary.
For example, a SCS should not be accused of being less like a macroscopic superposition
due to the reason that it does not look like a GHZ state, and vice versa.
This problem, together with the first one mentioned above, causes the previous measures
\cite{Leggett,Dur,Cavalcanti,Korsbakken,Bjork,Mar}
to be limited to specific types of superpositions, and/or
obscures comparisons between various types of states.
For example, it would be difficult to compare a GHZ state 
and a continuous-variable Gaussian state, and even worse when both the
states are somehow partially decohered.
In order to effectively compare different types of states 
in terms of their sizes as macroscopic superpositions,
a decomposition-independent (and measurement-independent) measure
that can be commonly applied to any  given state is highly desired.

In this Letter, we propose a novel measure 
that satisfies these requirements based on quantum interference
in phase space.
For an arbitrary given state in phase space, it provides quantitative
information about both the crucial aspects of
macroscopic superpositions: the effective size of the physical system that involves
the superposition and the degree of quantum coherence.
The appropriateness, inclusiveness and usefulness of our proposal are confirmed by 
(i)	its direct relation to a well-known decoherence model,
(ii) its direct relation to a previous measure \cite{Dur} proposed for a specific type of states,
(iii) its advantageousness in computability as a practical tool, and
(iv) various examples including mixed states with sensible results.

Among the previously proposed measures, Bj\"{o}rk \textit{et al.}'s one \cite{Bjork}
is based on interference between component states,
while it does not distinguish a pure superposition from a classical mixture. 
We attempt to consider quantum interference of a given state in a more general
framework using a phase space formalism.
Phase-space representations such as the Wigner function
are very useful 
to visualize a quantum state,
from which some crucial information can be intuitively obtained.
In terms of the Wigner function,
a macroscopic quantum superposition has two (or more) well-separate peaks
and has some oscillating patterns between them in phase space.
It is known that these interference fringes tend to appear more frequent
as the distinguishable peaks are more separate.
We pay attention to 
the ``frequency'' of the interference as an indicator of a macroscopic superposition.
Of course, it is a separate problem to quantify it in the phase-space structure
for a proper measure.

The characteristic function for a density operator $\rho$
for a single-mode case is defined as $\chi \left( \mathbf{\xi} \right)={\rm Tr}\{\rho \exp[\xi{\hat a}^\dagger-
\xi^*{\hat a}]\}$ where 
$\hat a$ and ${\hat a}^\dagger$ are the bosonic annihilation and creation operators, respectively.
The Wigner function $W (\alpha)$  
is the Fourier transform of the characteristic function \cite{QObook}
as
$W \!\left( \alpha_r, \alpha_i \right) =
\frac{1}{\pi^2} \int \!\! \int \! d \xi_r d \xi_i \; \chi \left( \xi_r, \xi_i \right) e^{-2 i (\alpha_r \xi_i - \alpha_i \xi_r)},$
where subscript $r ~ (i)$ denotes the real (imaginary) part of the given variable.
We notice that a frequency of a Wigner-function component along the real (imaginary) axis is
$\xi_i$ ($\xi_r$) and
its complex amplitude 
for specific frequency $\xi$ corresponds to $\chi \left( \mathbf{\xi} \right)$.

We know that
(i)	the frequency of the fringes (how dense the fringes are) reflects the ``effective size'' of the superposition
(i.e. how far the component states separate), and 
(ii) ``coherence'' (i.e. the degree of genuine superposition against its completely mixed version,
say, in terms of the ``pointer basis'' \cite{deco})
relates to the magnitude of the interference fringes.
In order to quantify both the features at the same time,
it is natural to take 
the sum over (size of frequency) $\times$ (absolute amplitude for the given frequency).
Here, we take it in the form as
$
\int \!\! d^2 \mathbf{\xi} \left( \xi_r^2 + \xi_i^2 \right)
\left| \chi \left( \mathbf{\xi} \right) \right|^2,
$
so that it quantifies both the ``frequency" and the ``magnitude" of interference fringes in the Wigner representation.

We present the formal definition of our interference-based measure as
\begin{eqnarray}
\label{Definition1}
\mathcal{I} \left( \rho \right)
=&& \frac{1}{2 \pi^M} \!\! \int \!\! d^2 \boldsymbol{\xi}
 \sum_{m = 1}^M \! \left[ \left| \xi_m \right|^2  - 1 \right] \left| 
 \chi \left( \boldsymbol{\xi} \right) \right|^2  \\
\label{Definition2}
= \frac{\pi^M}{2} &&\!\! \int \!\! d^2 \boldsymbol{\alpha} \; W \!\left(\boldsymbol{\alpha} \right)
 \sum_{m = 1}^M \! \left[ -\frac{\partial^2 }{\partial \alpha _m \partial \alpha _m^*} - 1 \right] W\! \left(\boldsymbol{\alpha} \right),
\end{eqnarray}
where $m$ indicates different modes, $M$ the number of such modes,
$\boldsymbol{\xi} = (\xi_1,\, \xi_2,\, \cdots,\, \xi_M)$, $\int d^2\boldsymbol{\xi} = \int d^2\xi_1 \int d^2\xi_2 \cdots \int d^2\xi_M$,
and $\boldsymbol{\alpha}$ and $\int d^2\boldsymbol{\alpha}$ are defined in the same manner.
The form of the above definition is based on several reasons that become clear in the remaining discussions.
Since the above definition is grounded on a general characteristic of the Wigner function,
it can be applied to any bosonic or multi-mode states not only for superpositions consisting of more than 
two components but also for any partially or fully mixed state.

A remarkable feature of $\mathcal{I} (\rho)$ is that it is directly related to a decoherence model as
\begin{equation}
\label{Definition3}
\mathcal{I} \left( \rho \right) =  - \mathrm{Tr} \left[ \rho \mathcal{L} \left( \rho  \right) \right],
\end{equation}
where $\mathcal{L} \left( \rho \right)$ is the superoperator in the Lindblad form of
a vacuum-environment decoherence model \cite{QObook,Louisell}:
\begin{equation}
\frac{d \rho}{d\tau}
=\mathcal{L} \left( \rho \right) = \sum_{m = 1}^M
\left[ \hat{a}_m \rho \hat{a}^\dagger_m - \frac{1}{2} \rho \hat{a}^\dagger_m \hat{a}_m - \frac{1}{2} \hat{a}^\dagger_m \hat{a}_m \rho \right],
\label{dmodel}
\end{equation}
where $\tau=$ (decay rate)$\times$(time) is the dimensionless time.
If we let $\mathcal{P} = \mathrm{Tr} \left( \rho^2 \right)$ be the purity of state $\rho$,
we find
\begin{equation}
\label{PurityDecay}
\frac{d \mathcal{P} \left( \rho \right)}{d\tau}  =
 -2 \; \mathcal{I} \left( \rho \right).
\end{equation}
Consequently,
$\mathcal{I} \left( \rho \right)$ can be interpreted as the decreasing rate of the purity of $\rho$.
This interpretation conforms exactly with one of D\"ur {\it et al}.'s~\cite{Dur}
where the authors suggested a size measure for a specific type of superpositions.
The form of superposition studied in Ref.~\cite{Dur} is
\begin{equation}
\label{DurExample}
\ket{\phi} ={\cal K}\left(\, \ket{\phi_1}^{\otimes N} + \ket{\phi_2}^{\otimes N} \right)
\end{equation}
with ${\cal K}$ being the normalization factor and $|\langle \phi_1 \ket{\phi_2}|^2 = 1- \epsilon^2 \neq 0$
with small real value $\epsilon$.
Following Ref.~\cite{Dur},
we take $\ket{\phi_1}=\ket{0}$ and $\ket{\phi_2}= \cos \epsilon \ket{0} + \sin \epsilon \ket{1}$
with assumptions $\epsilon^2 \ll 1$ and $N \epsilon^2 \gg 1$, and
we obtain $\mathcal{I} \left( \rho \right) \simeq N \epsilon^2 /4$ for state~(\ref{DurExample}).
Remarkably, this result is the same as the one in Ref.~\cite{Dur} only by a constant factor,
even though our measure is derived from  a  starting point quite different
from Ref.~\cite{Dur} where the effective particle number involving the superposition was concerned.

Along these lines, it is conjectured that even though our measure is based on the
size of the frequency of interference fringes, it is closely related to the number
of particles composing the superposition.
For example, it can simply be shown from Eq.~(\ref{Definition3}) that $\mathcal{I} \left( \rho \right)$ 
 exactly  gives the particle number $n$ for a bosonic number state $\ket{n}$.
Besides, 
$\mathcal{I} (\rho)$ properly assesses the degree of a true superposition
against incoherent mixtures.
The following theorem shows that only a pure state can give the maximum value of
$\mathcal{I} (\rho)$ for a given average particle number.
\emph{Theorem:}
 $\mathcal{I} \left( \rho \right)$
has the maximum value $\langle \hat{n} \rangle$,
the average number of particles for $\rho$, if and only if
$\rho$ is a pure state and is orthogonal to any one-particle-subtracted state of itself \cite{Proof}.
It follows that a mixed state has always a lower value of $\mathcal{I} \left( \rho \right)$ than its $\langle \hat{n} \rangle$.

It is straightforward to show that 
the maximum value of $\mathcal{I}(\rho)=\langle {\hat n}\rangle$ is obtained for
the SCS $\propto \ket{\alpha} + \ket{-\alpha}$, where $|\pm\alpha\rangle$ are coherent states of amplitudes $\pm\alpha$,
the GHZ state $\propto|0\rangle ^{\otimes N}+|1\rangle ^{\otimes N}$,
and the NOON state $\propto\ket{n}\!\ket{0}+\ket{0}\!\ket{n}$.
Note that the average particle number of the SCS is related to $\alpha$ as 
$\langle{\hat n}\rangle=\alpha^2\tanh\alpha^2$, where
$\alpha$ was assumed to be real without loss of generality.
On the other hand, as shown above, mixed versions of the aforementioned states
have values of $\mathcal{I} \left( \rho \right) < \langle \hat{n} \rangle$.
We note that $\mathcal{I} \left( \rho \right) = 0$
for fully mixed states such as 
$\frac{1}{d} I_{d \times d}$,
where $I_{d \times d}$ is a $d$-dimensional identity matrix,
and $\rho\propto \ket{\alpha} \bra{\alpha} + \ket{-\alpha} \bra{-\alpha}$.
This means that no matter how large the size of the system is,
if the state scarcely has potential for quantum interference, 
the measure $\mathcal{I} \left( \rho \right)$ gives the value close to zero.

\begin{figure}
{\scalebox{0.4}{\includegraphics{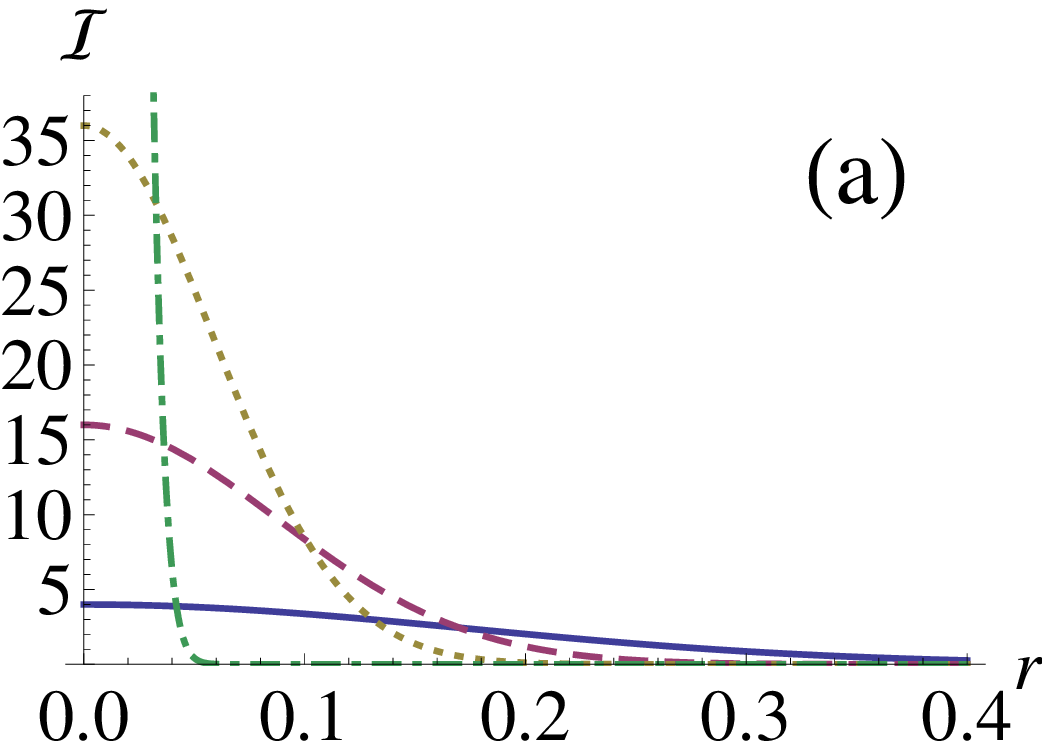}}
\scalebox{0.4}{\includegraphics{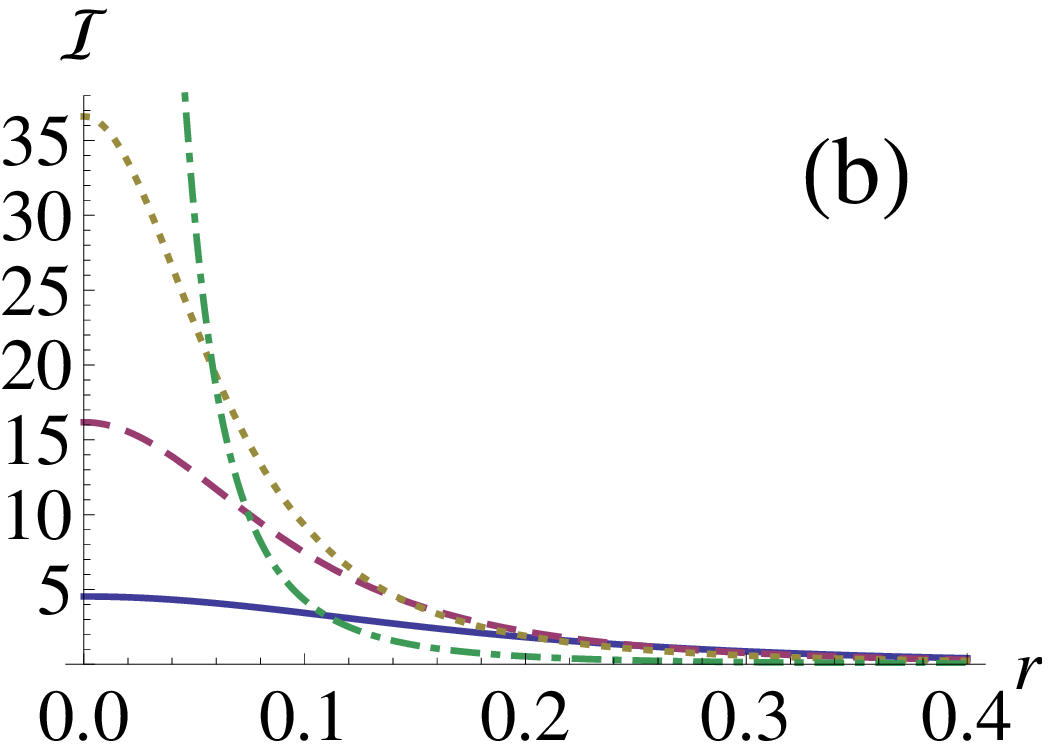}}}
\caption{(Color online) (a) Interference-based measure ${\cal I}(\rho)$ for SCSs of amplitudes $\alpha=2$ (solid curve), $\alpha=4$ (dashed), 
$\alpha=6$ (dotted) and $\alpha=27.3$ (dot-dashed) against the normalized time $r$ under the decoherence effect. The average number of particles is 
$\langle{\hat n}\rangle\approx\alpha^2$. (b) ${\cal I}(\rho)$ for a single-mode Gaussian states of squeezing parameters $s=1.5$ (solid), $s=2.1$ 
(dashed), $s=2.5$ (dotted) and $s=7$ (dot-dashed), where  $\langle{\hat n}\rangle=\sinh^2 r$. 
The same curve types mean (nearly) the same average particle numbers.
}
\label{figure1}
\end{figure}

Let us consider a partially mixed SCS under the decoherence effect caused by Eq.~(\ref{dmodel}): 
\begin{equation}
\rho={\cal N} \{\ket{t\alpha} \bra{t\alpha} + \ket{-t\alpha} \bra{-t\alpha}+
\Gamma(\ket{t\alpha} \bra{-t\alpha} + \ket{-t\alpha} \bra{t\alpha})\},
\end{equation}
where $t=e^{-\tau/2}$, $\Gamma=\exp[-2(1-e^{-\tau})\alpha^2]$, and $\cal N$ is
the normalization factor.
Using Eq.~(\ref{Definition1}), the interference-based measure is obtained as
$\mathcal{I}(\alpha,\tau)= 
\langle {\hat n}(0) \rangle
e^{-\tau} \sinh[2(2e^{-\tau}-1)\alpha^2]/\sinh [2\alpha^2]$
where $\langle {\hat n}(0) \rangle$ denotes  the average number of particles at $\tau=0$.
Here, the two crucial factors, 
the effective number of particles and the degree of true coherence,
are properly measured by $\alpha$ and $\tau$.
In Fig.~1(a), we plot ${\mathcal I}(\rho)$ for several cases of SCSs against the normalized time
$r=\sqrt{1-e^{-\tau}}$.
While SCSs with large amplitudes have large values of $\mathcal I(\rho)$,
they decrease more rapidly than SCSs with small amplitudes. 
This is satisfactorily in accordance with the well known fact, {\it i.e.},
the rapid destruction of macroscopic quantum superpositions \cite{deco,KB92}.
Here, a remarkable advantage of our measure is obvious that 
any fully or partially decohered superpositions are effectively quantified.

Our measure also provides sensible results for single-mode and multi-mode Gaussian continuous-variable
states which are useful for quantum information applications \cite{Kimble98}.
As an example, a general form of the characteristic function for a single-mode Gaussian state
is $\chi(\xi)=\exp[-A \xi_r^2/2-B \xi_i^2/2]$.
Real positive parameters $A$ and $B$ satisfy $AB\geq 1$ and the state is pure when the equality sign holds.
Using Eq.~(\ref{Definition1}), the interference-based measure is obtained as ${\cal I}(\rho)=(A+B-2AB)/[4 (AB)^{3/2}]$,  and
it is reduced to $(A+A^{-1}-2)/4$ for pure states. Obviously, the more ``squeezed'' pure state ($A\gg1$ or $A\approx0$) gives the larger value of 
${\cal I}(\rho)$,
and it approaches infinity in the limit of the original Einstein-Podosky-Rosen state
($A\rightarrow \infty$ or $A\rightarrow 0$). 
Suppose that a pure Gaussian state ($A=e^{-2s}$ and $B=e^{2s}$), where $s$ is the squeezing parameter,
is under decoherence described by Eq.~(\ref{dmodel}).
The time-dependent state is then characterized by $A=r^2+e^{-2s}t^2$ and $B=r^2+e^{2s}t^2$. The measure
${\cal I}(\rho)$ is immediately obtained from the result above, which has been plotted
in Fig.~1(b).
The results in Fig.~1(b) are qualitatively similar to those of SCSs in Fig.~1(a) 
for the same average particle numbers, while interestingly Gaussian states are more robust
against decoherence for large average particle numbers.

Our results with Gaussian states may not be very clear for first glance since
Gaussuan states do not show visible
interference patterns nor negativity in the Wigner function.
From our viewpoint, one reason for these reasonable results 
is that shrinking of the Wigner function into a narrow region in
phase space causes large frequency components to be dominant, 
which is a well known characteristic of the Fourier transform.

Recently, an exceptional type of macroscopic superpositions was introduced
from considering a more realistic analogy of Schr\"{o}dinger's
cat paradox \cite{JeongRalph2006}.
An example of such a state is 
$
\propto
\int d^2\alpha {\cal P}(V,d)
\{
|\alpha\rangle\langle\alpha|
+|-\alpha \rangle\langle\alpha|
+|\alpha\rangle\langle-\alpha|
+|-\alpha\rangle\langle-\alpha|
\}$
where 
${\cal P}(V,d)=
\exp[-\frac{2|\alpha-d|^2}{V-1}]$, $V$ is the variance of a thermal mixed-state component
and $d$ the distance between those components.
Such a state has prominent quantum properties but with large mixedness \cite{JeongRalph2006}.
Our measure also sensibly quantifies such a peculiar type of macroscopic superpositions
as 
\begin{equation}
\begin{aligned}
{\cal I}(V,d)={\cal M}^2& \Big[e^{-S}(Q-\frac{S}{V^2})+Q+S
\\&
-\frac{8e^{-\frac{V^2S}{U}}R\{RU-4d^2(V+1)\}}{U^3}\Big]
\end{aligned}
\label{mixresult}
\end{equation}
where ${\cal M}=(2+2V^{-1}e^{-\frac{S}{2}})^{-1}$,
$Q=(R/V)^2$, $R=V-1$, $S=4d^2/V$ and $U=V^2+1$.
Here, ${\cal I}(\rho)$ can be made arbitrarily large by increasing $d$ 
regardless of how large $V$ is.
The measure ${\cal I}(\rho)$ generally decreases by increasing $V$ when $d\gg0$. However,
when $d=0$, ${\cal I}(\rho)$ increases and saturates to a nonzero constant as ${\cal I}(\rho)\rightarrow 0.5$
  for $V\rightarrow\infty$, for which ``nonclassicality'' is effectively evidenced even though the Wigner function has neither negative part nor
squeezing properties. 
All these results with Eq.~(\ref{mixresult}) are perfectly in agreement with the tendency of the 
Bell inequality violations closely investigated with this type of states in Ref.~\cite{JeongRalph2006}.

Since the Fourier transform is invariant of any translation and rotation in the integration region,
$\mathcal{I} \left( \rho \right)$ is also invariant of any translation and rotation in phase space, {\it i.e.},
$
\mathcal{I} \left( U \rho U^\dagger \right) = \mathcal{I} \left( \rho  \right)
$,
where $U$ is such a translation or rotation.
This property frustrates certain attempts to ``artificially'' increase $\mathcal{I}(\rho)$
by adding particles to the systems.
For exampe, a coherent state $\ket{\alpha}$ displaced from the vacuum state $\ket{0}$ has the same value of $\mathcal{I}=0$ as $\ket{0}$. 
It is worth noting that 
 having no preferred basis states
is implied as one of the necessary conditions of a measure being a faithful size criterion
for macroscopic superpositions in 
Ref.~\cite{Bjork}
and it seems that our measure satisfies such requirement.

We point out that $\mathcal{I} (\rho)$
does not need any asymptotic assumptions or optimization techniques
as in Refs.~\cite{Cavalcanti,Korsbakken}.
From a practical point of view,
it is very simple to calculate using any of Eqs.~(\ref{Definition1})-(\ref{Definition3})
for an arbitrary quantum state.
Especially,
even for an experimentally generated state,
$\mathcal{I} (\rho)$ can be evaluated
using definition (\ref{Definition2}) based on the Wigner function
reconstructed by the tomography technique, {\it i.e.},
without the help of the fidelity with respect to a target state,
its quality can readily be assessed.

To extend our proposal to atomic/spin systems  
represented in a finite-demensional Hilbert space, one needs to apply the
discrete Wigner function and Fourier transform.
In this case, the corresponding master equation is also replaced
with an appropriate one considering the underlying Hilbert space.
We finally note that our measure does not suggest a
threshold beyond which a superposition is ``macroscopic'' 
but rather it provides a continuous scale to compare sizes 
of different superpositions.

In summary, we have proposed a measure  
to quantify macroscopic quantum superpositions. 
Using our measure, true quantum coherence and the effective size of the system 
that involves the superposition are simultateously quantified.
Interestingly, it is directly connected to a well known decoherence model and 
corresponds to the decay rate of the purity for the given state.
It has been found from this relevance that our general measure is in 
accordance with D\"{u}r \textit{et al.}'s designed for a specific type of states \cite{Dur}.
Since $\mathcal{I} (\rho)$ is based on the Wigner representation,
which completely describes a quantum state,
it is decomposition-independent 
and easy to calculate for any states represented in phase space including mixed states, giving
definite values for direct comparison between different types of states.
All these features are hardly seen in previously proposed measures.
Our measure will be widely useful for theoretical and experimental 
studies on macroscopic quantum systems and various related issues.

\acknowledgments
This work was supported by the NRF grant funded by
the Korea government (MEST) (No. 3348-20100018) 
and the WCU program.
H.J. acknowledges support from TJ Park Foundation.


\begin{thebibliography}{99}
\bibitem{Proof}
The following proof is for the single-mode case
but simply extended to multi-modes.
When evaluating $\mathrm{Tr} (\rho \sigma)$ for positive matrices $\rho$ and $\sigma$,
one can assume without loss of generality that
$\rho$ is a diagonal matrix
since $\rho$ can be diagonalized as
$\rho = U^\dagger \tilde{\rho} U$,
$\mathrm{Tr} (\rho \sigma) = \mathrm{Tr} (\tilde{\rho} \, U \sigma U^\dagger)$
and then $U \sigma U^\dagger$ is also a positive matrix.
Therefore,
since the diagonal elements $\sigma_{ii} \geq 0 $ for a positive matrix $\sigma$,
we have
$\mathrm{Tr} (\rho \sigma)
= \sum_{i} \rho_{ii} \sigma_{ii}
\leq \rho_{ii}^{\textrm{max}} \sum_{i} \sigma_{ii}
\leq \mathrm{Tr} (\sigma),
$
where the equality sign holds when $\rho_{ii}^{\textrm{max}}=1$, that is, $\rho$ is a pure state.
Substituting
$\sigma =
\left( \rho \hat{a}^\dagger \hat{a} + \hat{a}^\dagger \hat{a} \rho \right)/2$
in the above relation, we get
$
\mathrm{Tr} \left[ \rho \, 
\left( \rho \hat{a}^\dagger \hat{a} + \hat{a}^\dagger \hat{a} \rho \right) \right/2]
\leq \mathrm{Tr} (\rho \hat{a}^\dagger \hat{a})
= \langle \hat{n} \rangle.
$
Similarly, since
$
\mathrm{Tr} (\rho \sigma) \geq \rho_{ii}^{\textrm{min}} \sum_{i} \sigma_{ii}
\geq 0 \cdot \mathrm{Tr} (\sigma) = 0,
$
we also get
$
\mathrm{Tr} (\rho \hat{a} \rho \hat{a}^\dagger )
\geq 0,
$
and finally
$\mathrm{Tr} [\rho \mathcal{L} (\rho) ]
\leq \langle \hat{n} \rangle - 0
= \langle \hat{n} \rangle,
$
where the equality sign holds if and only if $\rho$ is a pure state and
$\mathrm{Tr} (\rho \, \hat{a} \rho \hat{a}^\dagger) = 0$.
\end{thebibliography}
\end{document}